\documentstyle[12pt]{article}
\begin{document}
\par
\vspace{10mm}
\centerline{\LARGE{\bf{Global monopole in scalar tensor theory}}}
\vspace{10mm}
\centerline{\bf{A.Banerjee$^{1}$,A.Beesham$^{3}$,S.Chatterjee$^{1,2}$,A.A.Sen$^{1}$}}
\vspace{5mm}
\centerline{$^{1}$Relativity and Cosmology Research Centre}
\centerline{Department of Physics,Jadavpur University}
\centerline{Calcutta-700032,India.}
\par
\vspace{5mm}
\begin{center}
$^{2}$Department of Physics\\
New Alipore Govt.College\\
Calcutta-700053,India.\\
\vspace{5mm}
$^{3}$Department of Applied Mathematics\\
University of Zululand\\
Kwadlanguzwa 3886\\
South Africa.\\
\end{center}
\par
\vspace{5mm}
\par
\vspace{5mm}
PACS NO:04.20.Jb,04.25.Dm,04.50.+h.           \\      
\vspace{5mm}

Correspondence To : A.A.Sen.\hspace{4mm}E-mail :anjan@juphys.ernet.in                   \\

\newpage        
\par
\vspace{10mm}
\underline{\bf{Abstract}}
\vspace{5mm}
\par
The well known monopole solution of Barriola and Vilenkin (BV) resulting from the 
breaking of a global SO(3) symmetry is extended in general relativity along
with a zero mass scalar field and also in Brans-Dicke(BD) theory of gravity.In the case
of BD theory , the behaviour of spacetime and other variables such as BD scalar
field and the monopole energy density have been studied numerically.For monopole
along with a zero mass scalar field , exact solutions are obtained and depending
upon the choice of arbitary parameters , the solutions either reduce to the BV case 
or to a pure scalar field solution as special cases.It is interesting to note that unlike the BV case
the global monopole in the BD theory does exert gravitational pull on a test particle
moving in its spacetime.\\
\par
\par
\par

\newpage
\section{Introduction}
\par
Monopoles are point like topological objects that may arise during phase 
transitions in the early universe\cite{R1,R2}.Depending on the nature of the scalar 
field it can be shown that spontaneous symmetry breaking can give rise to such 
onjects which are nothing but the topological knots in the vacuum expectation 
value of the scalar field and most of their energy is concentrated in a small region
near the monopole core.From the topological point of view they are formed in 
vacuum manifold when the latter contains surfaces which can not be continuously
shrunk to a point.These monopoles have Goldstone fields with their energy 
density decreasing with distance as $r^{-2}$.These monopoles are found to have some
interesting features in a sense that it exerts practically no gravitational
force on its surrounding nonrelativistic matter but the spacetime around it
has a deficit solid angle.
\par
In a pioneering work Barriola and Vilenkin(BV) \cite{R3} showed the existence 
of such a monopole solution resulting from the breaking of global SO(3) symmetry
of a triplet scalar field in a Schwarzschild background.Recently we \cite{R4}
have also obtained a monopole solution in Kaluza-Klien spacetime which extends the
earlier work of BV to its five dimensional analogue.
\par
In the present work we make an attempt to study such a monopole in Brans-Dicke(BD)
theory\cite{R5} of gravitation .The renewed interest of scalar tensor theory 
of gravitation is mainly due to two important theoretical developments in the 
study of early universe - one is the prediction of the dilaton field arising from
the low energy limit of the string theory.The other is the recent theory of extended 
inflation which is beleived to have solved the fine tuning problem by slowing 
down the expansion rate of universe from the exponential to polynomial.
\par
In view of serious difficulties in solving the monopole problem in the original 
version of the BD theory,an attempt is made in the second part of the paper
to get the solution in the Dicke's revised unit\cite{R6} where the field equations
are identical with the Einstein's field equations with the BD scalar field 
appearing as a "matter field" in the theory.But even in the revised version
,we have not as yet obtained the solution in a closed form.However we have been
able , so far , to reduce the mathematical formalism to a single differential
equation involving only a single variable namely the BD scalar field and once the
scalar field is known , the metric components and the monopole energy density
can be obtained from the other equations.We have,however,made a numerical
study of the behaviour of the different variables and the results seem to be 
physically consistent.
\par
In the third part of the paper we have studied the similar problem with the monopole
interacting with a zero mass scalar field and have been able to obtain the
exact solutions of the field equations.The BV monopole solutions are recovered
from these solutions when the zero mass scalar field vanishes.On the other hand
when the monopole field is switched off we get back the usual zero mass scalar
field solution already existing in literature.
\par
Our paper is organised as follows. In section 2 the field equations for a 
global monopole in Brans-Dicke theory in revised units are written, analysed, 
and their qualitative behaviour is studied numerically. In section 3 the same
problem is addressed in Einstein's theory in the presence of a zero mass scalar
field for two distinctly different situations. The paper ends with a conclusion
in section 4.
\par
\par
\par
\section{The Global Monopole In Branse Dicke Theory}
\par
The gravitational field equation for a global monopole in Brans-Dicke theory
written in Dicke's revised unit\cite{R6} is in general
$$
G^{\mu}_{\nu} = - T^{\mu}_{\nu} - {{2\omega+3}\over{2}}{1\over{\phi^{2}}}[\phi^{,\mu}\phi_{,\nu}
-{1\over{2}}\delta^{\mu}_{\nu}\phi_{,\alpha}\phi^{,\alpha}]
\eqno{(1)}$$
\par
Where $T^{\mu}_{\nu}$ is the energy momentum tensor due to a monopole field,
$\phi$ is the B-D scalar field and $\omega$ is the B-D parameter.
\par
Since the spacetime here is static and spherically symmetric, the metric is 
given in curvature coordinates in the form
$$
ds^{2} = e^{\nu}dt^{2}-e^{\beta}dr^{2}-r^{2}dr^{2}-r^{2}sin^{2}\theta d\Phi^{2}
\eqno{(2)}$$
\par
where $\nu$ and $\beta$ are functions of r alone.
\par
In the original work of Barriola and Vilenkin\cite{R3} there is no contribution
to the energy momentum tensor from  the B-D scalar field but the lagrangian is 
due to global SO(3) symmetry for a triplet scalar field, whose symmetry
breaking gives rise to a global monopole. Under reasonable conditions the 
energy momentum tensor due to the monopole field outside the core now becomes
\cite{R3}
$$
T^{t}_{t} = T^{r}_{r} = {\eta^{2}\over{r^{2}}}
\eqno{(3)}$$
$$
T^{\theta}_{\theta} = T^{\Phi}_{\Phi} = 0
\eqno{(4)}$$
\par
where $\eta$ is the symmetry breaking scale of the theory.
\par
The above forms of $T^{\mu}_{\nu}$ are consistent with the Bianchi identity.
One should note here that in the original version of the B-D theory\cite{R5}
the relation $T^{\mu}_{\nu;\mu}=0$ is separately satisfied which again leads
to the same expressions (3) and (4) for $T^{\mu}_{\nu}$. However, in the 
revised units, the B-D scalar field has additional contributions to the 
energy momentum tensor in the field equations and consequently the Bianchi 
identity in this case will certainly not lead to the identical expressions for
energy momentum tensors as for monopole field only. We therefore prefer to write

$$
T^{t}_{t} = T^{r}_{r} = \rho(r)
\eqno{(5)}$$
$$
T^{\theta}_{\theta} = T^{\Phi}_{\Phi} = 0 
\eqno{(6)}$$
\par
where $\rho(r)$ is a function of r to be determined from the field equations.
\par
With (2),(5) and (6),  equation (1) yields explicitly the following relations:
$$
e^{-\beta}({1\over{r^{2}}}-{\beta^{'}\over{r}})-{1\over{r^{2}}}=-\rho-{k\over{2}}\psi^{'2}e^{-\beta} 
\eqno{(7a)}$$
$$
e^{-\beta}({1\over{r^{2}}}+{\nu^{'}\over{r}})-{1\over{r^{2}}}=-\rho+{k\over{2}}\psi^{'2}e^{-\beta}   
\eqno{(7b)}$$
$$
e^{-\beta}({\nu^{''}\over{2}}+{\nu^{'2}\over{4}}-{{\nu^{'}\beta^{'}}\over{4}}+{{\nu^{'}-\beta^{'}}\over{2r}})=-{k\over{2}}\psi^{'2}e^{-\beta} 
\eqno{(7c)}$$
\par
where $\psi=ln(\phi)$ and $k={{2\omega+3}\over{2}}$ and prime denotes differentiation
with respect to r.
\par
The wave equation for the BD scalar field is given by
$$
\psi^{''}+2{\psi^{'}\over{r}}+({{\nu^{'} -\beta^{'}}\over{2}})\psi^{'} = - \rho{e^{\beta}\over{k}}
\eqno{(8)}$$
By (7a)-(7b)-2x(7c) one gets
$$
\nu^{''}+2{\nu^{'}\over{r}}+{\nu^{'2}\over{2}}={{\nu^{'}\beta^{'}}\over{2}}
\eqno{(9)}$$
\par
One may attempt to solve the eqn(9) for two different situations:
(i)$\nu^{'}=0$, which means $\nu=constant$ and (ii)$\nu^{'}\neq 0$, which 
in turn leads to differential equation: $e^{\nu}\nu^{'2}={b^{2}\over{r^{4}}}e^{\beta}$,
where $b$ is an arbitrary constant of integration.
\par
In the first case one may choose $\nu=1$ without the loss of any generality,
which in view of Bianchi identity relation immediately leads to the equation
$$
\rho^{'}+2{\rho\over{r}}+\rho\psi^{'}=0
$$
The above equation has the first integral
$$
\rho r^{2}=a^{2}e^{-\psi}
\eqno{(10)}
$$
where $a$ is another arbitrary integration constant.
\par
In the second case that is for $\nu^{'}\neq 0$ it is extremely hard task to 
obtain any differntial equation which involves only a single variable. Hence
we abandon this case in our present work and proceed with the other case, 
that is, $\nu^{'}=0$ for further calculations.
\par
With $\nu^{'}=0$ one gets after some simple and straightforward calculation
the following two equations:
$$
e^{\beta}={{({k\over{2}}\psi^{'2}r^{2}-1)}\over{(a^{2}e^{-\psi}-1)}}
\eqno{(11)}$$
$$
(a^{2}-e^{\psi})\psi^{''}-{k\over{2}}r(a^{2}-e^{\psi})\psi^{'3}+{a^{2}\over{2}}\psi^{'2}+
{2\over{r}}(a^{2}-e^{\psi})\psi^{'}-{a^{2}\over{kr^{2}}}=0
\eqno{(12)}$$
One should note that the actual BD scalar field is $\phi=e^{\psi}$.
\par
The equation(12) is the key equation to our subsquent analysis.If this equation can
be integrated,the complete set of equations will be solvable,as in view of equation(10)
and (11),$\rho$ and $e^{\beta}$ may be known in terms of $\psi$.
However equation(12) is a highly nonlinear equation which may not be amenable to an analytic
solution in a closed form.
\par
Hence we try to solve the equation (12) numerically in order to  
estimate the behaviour of scalar field $\phi,e^{\beta}$ as well as the energy density
of the monopole $\rho$ outside the core.
We rewrite the equation(12) as,
$$
{{{d^2}\phi}\over{dx^{2}}}=2.304{d\phi\over{dx}}+{{(a^{2}-2\phi)}\over{(2a^{2}-2\phi)\phi}}({d\phi\over{dx}})^{2}-
{0.434k\over{2\phi^{2}}}({d\phi\over{dx}})^{3}+{a^{2}\phi\over{0.434k(a^{2}-\phi)}}
\eqno{(13)}$$
where $x=log_{10}({\delta\over{r}})$ where $\delta$ = the core radius of the monopole,
$\phi=e^{\psi}$ , the BD scalar field.
The variables $e^{\beta}$ and $\rho$ take the form
$$
e^{\beta}={{0.095k({d\phi\over{dx}})^{2}-\phi^{2}}\over{a^{2}\phi-\phi}}
\eqno{(14)}$$
$$
\rho={{{a^2}e^{4.608x}}\over{\phi}}
\eqno{(15)}$$
\par
We have tried to solve the equation(13) numerically by fourth order Runge-Kutta 
method and plot $\phi,\phi^{'},\rho ,e^{\beta}$ against $r$.\\
We have taken the initial condition that for a large distance from the monopole core
i.e when $r$ takes a very large value, $\phi^{'} \rightarrow 0$ and $\phi$ approaches a very 
small value.We have taken the distance upto $10^{5}$ times the core radius $\delta$
of the monopole.
\par
In figure(1) we have plotted  both $\phi$ and $\phi^{'}$ and further in figure(2) and(3)
we plot $e^{\beta},\rho$ respectively against $log_{10}(\delta/r)$.
\par
The behaviour of $e^{\beta}$ in fig(2) shows that the spacetime outside the monopole
core becomes flat for a large distance from the core.The fig(3) shows that 
the monopole energy density $\rho$ rapidly decreases with respect to the distance
from the core.This is qualitatively consistent with the corresponding BV case where also 
the energy density of the monopole outside the core falls off with $r$.The
behaviour of scalar field in fig(1) also shows that $\phi$ approaches a constant 
value and $\phi^{'}$ approaches zero for a large distance from the core.
\par
\par
\par
\par
\vspace{5mm}
\centerline{\bf{Gravitational force in the field of a monopole}}
Before ending up  this section we would like to point out an important property
of the BD scalar field-the gravitational force acting on surrounding matter
of the monopole.
\par
In the original version of BD theory our metric (2) becomes 
$$ 
ds^{2} = e^{-\psi}dt^{2}-e^{-\psi}(e^{\beta}dr^{2}+r^{2}d\theta^{2}+r^{2}sin^{2}\theta d\Phi^{2})
\eqno{(16)}$$
as $\nu=constant $ for our case and $\psi$ , $\beta$ are given by (13) and (14)
respectively. The radial component of the acceleration acting on a test particle in the 
gravitational field  of the monopole is given by:
$$
\dot{v^{1}}=v^{1}_{;0}v^{0}
\eqno{(17)}$$
\par
Since for a comoving particle $v^{\mu}={1\over{\sqrt{g_{00}}}}\delta^{\mu}_{0}$
we have $v^{0}=e^{\psi/2}$ and $\dot{v^{1}}=-{1\over{2}}e^{(\psi-\beta)}\psi^{'}$.
Assuming $\psi^{'}<0$ which is obvious from fig(1),
the particle accelarates away in the radial
direction in order to keep it at rest which implies that a monopole has an attractive
influence on the test particle.
This attractive force is due to the presence
of the BD scalar field $\psi$.When $\psi^{'}=0$  the BD scalar field is absent and
we have only the monopole remaining which however exerts no gravitational influence
on the surrounding matter as shown by BV\cite{R3}.
\par
\section{Global monopole interacting with a zero mass scalar field}
\par
In this section we have considered a monopole in the presence of a massless
scalar field which also contribute to the energy momentum tensor in addition
to that of the monopole field. Exact analytical solutions may be obtained in 
this case. However, for economy of space the details of mathematical steps are
omitted and the main results are sketched below:\\
\underline{Case(i)}
\par
In this case we take the metric as 
$$
ds^{2}=e^{\nu}dt^{2}-e^{\beta}dr^{2}-r^{2}d\Omega^{2}
\eqno{(18)}$$
where $\nu$,$\beta$ are functions of $r$ only.\\
Solving the field equations we get 
$$
e^{\nu}=(1-\eta^{2}-2{M\over{r}})^{n}
\eqno{(19)}$$
$$
e^{\beta}=(1-\eta^{2}-2{M\over{r}})^{n-2}
\eqno{(20)}$$
Where $M$ and $n$ are arbitary constant of integration and $\eta$ is the symmetry
breaking scale for the monopole.\\
The energy density of the monopole is given by
$$
\rho_{mon}={1\over{r^{2}}}[1-(1-\eta^{2})(1-\eta^{2}-2{M\over{r}})^{n-1}]
\eqno{(21)}$$
The zero mass scalar field is given by
$$
\psi^{'2}={{4M(n-1)}\over{(1-\eta^{2})r^{3}}}(1-\eta^{2}-2{M\over{r}})^{-1}
\eqno{(22)}$$
For $n=1$ the variation of the scalar field $\psi$ vanishes,and we get back
the BV type monopole with $\rho_{mon}={\eta^{2}\over{r^{2}}}$\\
\par
On the other hand pure scalar field solution without the presence of a monopole
field cannot be recovered in this form of metric, as is evident from the expressions
(21) and (22).\\
\underline{Case(ii)}
\par
Here we take the metric in the form 
$$
ds^{2}=e^{\nu}dt^{2}-e^{\beta}dr^{2}-r^{2}R^{2}d\Omega^{2}
\eqno{(23)}$$
where $\nu,\beta$ and $R$ are functions of $r$ only.\\
The solutions are
$$
e^{\nu}=(1-\eta^{2}-2{M\over{r}})^{n}
\eqno{(24a)}$$
$$
e^{\beta}=(1-\eta^{2}-2{M\over{r}})^{-n}  
\eqno{(24b)}$$
$$
R^{2}=(1-\eta^{2}-2{M\over{r}})^{1-n}  
\eqno{(24c)}$$
$$
\rho_{mon}={\eta^{2}\over{r^{2}}}(1-\eta^{2}-2{M\over{r}})^{n-1}  
\eqno{(24d)}$$
$$
\psi^{'}=\sqrt{2}{M\over{r^{2}}}\sqrt{n^{2}-1}(1-\eta^{2}-2{M\over{r}})^{-1}  
\eqno{(24e)}$$
\par
It is clear from the expressions (24d) and (24e) that when $n=1$, we have
$\rho_{mon}=\eta^{2}/r^{2}$ and $\psi^{'}=0$, so that the BV solution is 
recovered. But on the other hand when we put $\eta^{2}=0$, we obtain
$\rho_{mon}=0$ and $\psi^{'}\neq 0$ meaning that we have a pure massless
scalar field distribution.
\par
\section{Conclusion}
\par
In this work we have extended the monopole solution of BV to the scalar tensor theory.
In BD theory we have not been able to get the analytical solution.So we have taken 
recourse to numerical methods to study the behaviour of BD scalar field.From this 
field ,the behaviour of metric coefficient and the energy density of the monopole
are also found out.It is encouraging to point out that the nature of variation of 
BD scalar field,metric coefficient and also the energy density of the monopole are quite
physically realistic in the sense that the spacetime is asymptotically flat.Also 
the monopole energy density outside the core decreases sharply with the distances.
\par
For the sake of completeness ,we have also studied the similar problem with monopole interacting with a massless 
scalar field,and have obtained exact solutions in this case.Under suitable choice of the 
parameters  in some cases our solutions reduce to either the BV solution or to that of a pure zero mass scalar 
field.
\par
To end up, a final remark may be in order.It is well known that in Einstein's
theory,the BV type of monopole does not practically exert any gravitational
pull on any surrounding matter.However our analysis shows that the monopole in BD
theory does exert gravitational force on a test particle.So this observation
is in striking contrast with the analogue in Einstein's theory.
\par
\par
\vspace{10mm}
\par
\section{Acknowledgement}
\par
S.C wishes to thank the University of Zululand,South Africa,for local hospitability
where a part of the work was done.A.A.S.thanks the University Grants Commission,
India for financial support.A.B. and S.C. also thank D.S.T.,India,for financial support.\\
\newpage
\par
\vspace{10mm}
\centerline{\bf{Figure Captions}}
\par
1>In Fig1 , we have plotted the BD scalar field $\phi$ and its first derivative  $\phi^{'}$
 with respect to $log(\delta/r)$.
\par
2>In Fig2 , we have plotted the metric coefficient $e^{\beta}$ with respect to
$log(\delta/r)$.
\par
3>In Fig3 , we have plotted the monopole energy density $\rho$ with respect 
to $log(\delta/r)$.\\

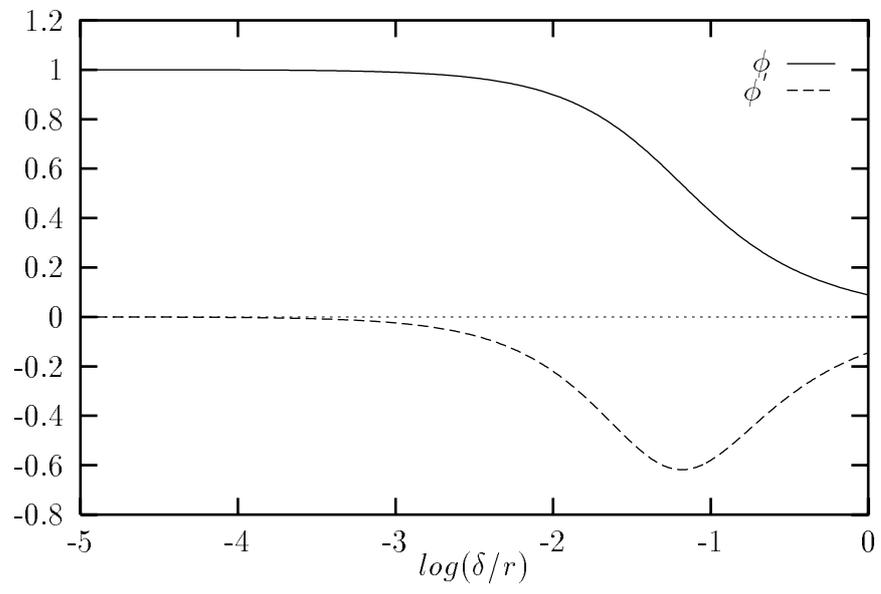
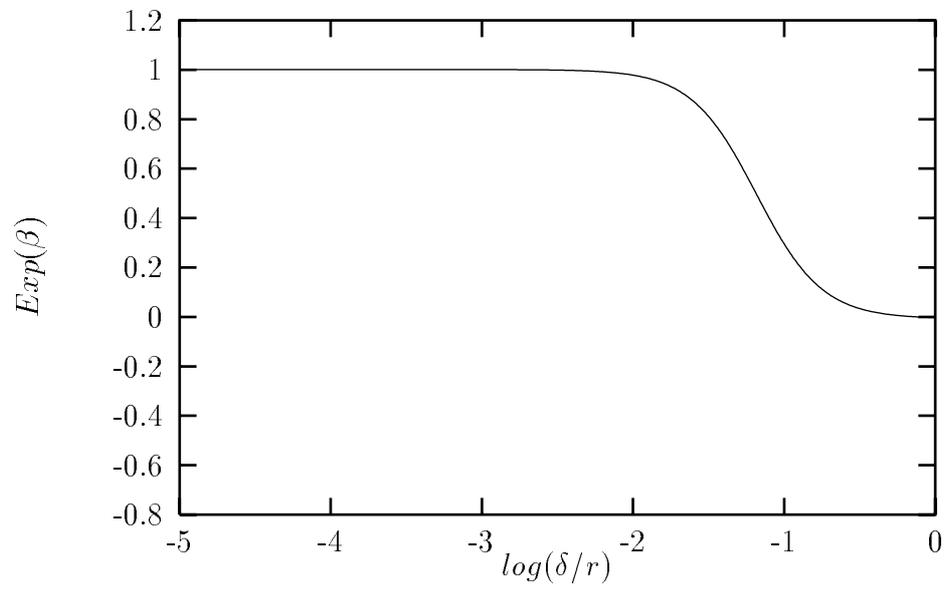
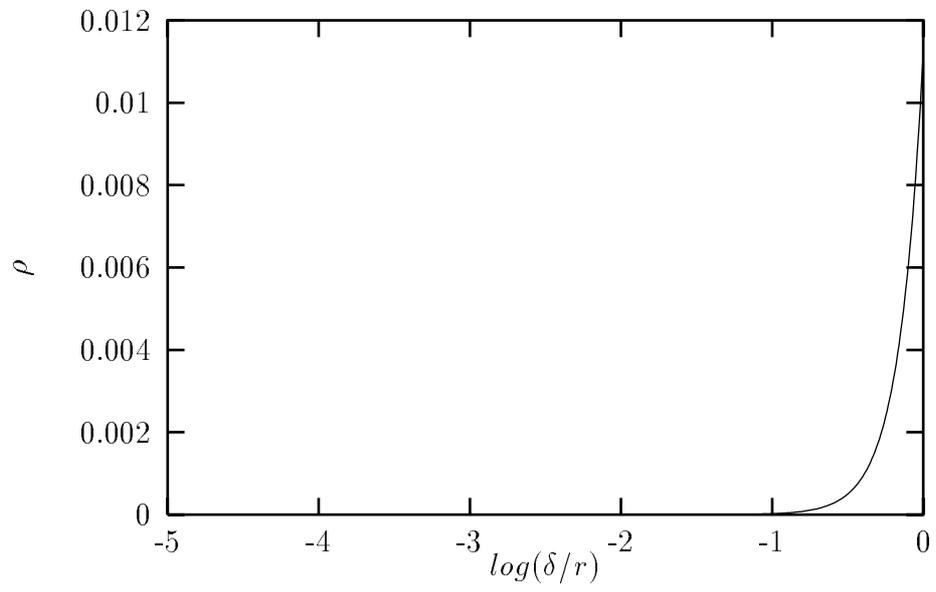
\end{document}